\documentclass[journal]{IEEEtran}
\usepackage[latin1]{inputenc}
\usepackage{times,amsmath}
\usepackage{amssymb}
\usepackage{steinmetz}
\usepackage{pstool}
\usepackage{subfigure}
\usepackage{multirow}
\usepackage{enumerate}
\usepackage{graphicx}
\usepackage{multirow}
\usepackage{subfig}
\usepackage{mwe}
\usepackage{bigints}
\usepackage{MnSymbol}
\usepackage{stfloats} 
\usepackage[table]{xcolor}
\usepackage[square, comma, sort&compress, numbers]{natbib}
\usepackage{nohyperref}
\usepackage{algorithm,algorithmic}
\usepackage{epstopdf}
\usepackage{mathtools, cuted}

\graphicspath{ {Figures/} }

\begin{document}

\title{ Non-Contact Monitoring of Dehydration using RF Data Collected off the Chest and the Hand}

\author{
\IEEEauthorblockN{
Hasan Mujtaba Buttar\IEEEauthorrefmark{1}, Kawish Pervez\IEEEauthorrefmark{1}, M. Mahboob Ur Rahman\IEEEauthorrefmark{1}, Kashif Riaz\IEEEauthorrefmark{1}, Qammer H. Abbasi\IEEEauthorrefmark{2} }

\IEEEauthorblockA{\IEEEauthorrefmark{1} Electrical engineering department, Information Technology University, Lahore 54000, Pakistan\\ 
\IEEEauthorrefmark{2}Department of Electronics and Nano Engineering, University of Glasgow, Glasgow, G12 8QQ, UK\\
\IEEEauthorrefmark{1}\{phdee21006, mahboob.rahman\}@itu.edu.pk, \IEEEauthorrefmark{2}Qammer.Abbasi@glasgow.ac.uk }
}

\maketitle

\begin{abstract} 
In this work, we report for the first time a novel non-contact method for dehydration monitoring from a distance. Specifically, the proposed setup consists of a transmit software defined radio (SDR) that impinges a wideband radio frequency (RF) signal (of frequency 5.23 GHz) in the microwave band onto either the chest or the hand of a subject who sits nearby. Further, another SDR in the closed vicinity collects the RF signals reflected off the chest (or passed through the hand) of the subject. Note that the two SDRs exchange orthogonal frequency division multiplexing (OFDM) signal, whose individual subcarriers get modulated once it reflects off (passes through) the chest (the hand) of the subject. This way, the signal collected by the receive SDR consists of channel frequency response (CFR) that captures the variation in the blood osmolality due to dehydration. The received raw CFR data is then passed through a handful of machine learning (ML) classifiers which once trained, output the classification result (i.e., whether a subject is hydrated or dehydrated). For the purpose of training our ML classifiers, we have constructed our custom HCDDM-RF-5 dataset by collecting data from 5 Muslim subjects (before and after sunset) who were fasting during the month of Ramadan. Specifically, we have implemented and tested the following ML classifiers (and their variants): K-nearest neighbour (KNN), support vector machine (SVM), decision tree (DT), ensemble classifier, and neural network classifier. Among all the classifiers, the neural network classifier acheived the best classification accuracy, i.e., an accuracy of 93.8\% for the proposed chest-based method, and an accuracy of 96.15\% for the proposed hand-based method. Compared to the state-of-the-art (i.e., the contact-based dehydration monitoring method) where the reported accuracy is 97.83\%, our proposed non-contact method is slightly inferior (as we report a maximum accuracy of 96.15\%); nevertheless, the advantages of our non-contact dehydration method speak for themselves. That is, our proposed method is non-invasive and contact-less, has high accuracy, allows continuous and seamless monitoring, is easy to use, and provides rapid results. The anticipated beneficiaries of the proposed method include: sportsmen, athletes, elderly, diabetic and diarrhea patients, and labor working outdoors.

\end{abstract}

\section{Introduction}
\label{sec:intro}

A good sixty percent of the human body is composed of water, which is essential to many of the body's activities, including maintaining the body's temperature, transporting nutrients and oxygen to cells, lubricating joints, and eliminating waste products. Consuming sufficient water on a daily basis is necessary for preserving one's health and warding off a variety of diseases and adverse conditions \cite{popkin2010water}. Dehydration occurs when the body does not obtain enough water or when the body loses water through sweating and evaporation. When dehydration occurs, it throws off the natural equilibrium of the minerals and electrolytes found in the body. This could result in a variety of different health issues, ranging from quite harmless to life-threatening, depending on how much fluid is lost and what's causing it in the first place. Symptoms of mild dehydration include headache, dry mouth, thirst, dizziness, exhaustion, and dry and wrinkled skin \cite{el2015acute},\cite{liaqat2021review},\cite{black1944study}. 
In more extreme circumstances, dehydration can result in consequences such as kidney failure, convulsions, and even death. When the outside weather is hot and humid, then the dehydration could lead to heat exhaustion which could induce symptoms such as heavy perspiration, nausea, headache, and weakness. Heat exhaustion if not addressed quickly, could escalate to heatstroke, which is a life-threatening medical emergency that can cause damage to the brain, organ failure, and even death. Last but not the least, dehydration could also have some long-term adverse effects on the body, e.g., constipation, damage to the kidneys, and infections of the urinary tract, etc. \cite{popkin2010water}.

In short, dehydration could have fatal implications if left untreated, thus, timely diagnosis of dehydration followed by imminent medical intervention is of utmost importance. For the elderly, and for the diabetic and diarrhea patients, it is especially important to track their hydration levels frequently \cite{el2015hydration}. But when it comes to the existing dehydration detection methods, they have their limitations as they are either invasive (e.g., blood sample based), or contact-based (e.g., pulse oximeter, smart watch based). Further, the existing methods are expensive, inconvenient and inconsistent, as discussed below.

{\bf Existing dehydration measures and the dilemma:}
Some of the most common methods for measuring hydration levels are: body mass change, total body water, serum and urine osmolality, plasma osmolality, urine specific gravity, and urine volume \cite{armstrong2007assessing,armstrong1994urinary,cheuvront2010biological,oppliger2005accuracy,popowski2001blood}.
Another method that is sometimes considered as the "gold standard" consists of a procedure whereby a subject ingests a known quantity of an isotope, which allows one to calculate its concentration in a bodily fluid in order to determine the body's total water content. Now, the dilemma. Though such "gold standards" of hydration assessment are considered useful for sports science, medicine, or for creating reference standards, but since they necessitate extensive methodological control, they are not useful for tracking one's hydration status on daily basis during a training or competition \cite{cheuvront2006hydration}. In other words, none of aforementioned hydration measures has been demonstrated to be valid in all dehydration scenarios (i.e., lab and field) \cite{kulkarni2020monitoring}. Last but not the least, many of the aforementioned hydration measures could be expensive, cumbersome, erroneous, and inconvenient (either invasive or contact-based). This calls for the innovative and preferably non-contact methods for dehydration monitoring, which is precisely the agenda of this work.

{\bf Contributions.} 
This paper proposes an RF-based dehydration monitoring method that is non-invasive and contact-less, has high accuracy, allows continuous and seamless monitoring, is easy to use, and provides rapid results.
Specifically, the key contributions of this work are as follows:

1) We propose a novel non-contact method called chest-based dehydration monitoring (CBDM) method. Under this method, the subject sits nearby an RF transceiver that impinges an OFDM signal onto the chest of the subject, while the receiver collects the signal reflected off the chest of the subject. 

2) We propose a novel non-contact method called hand-based dehydration monitoring (HBDM) method. Under this method, the subject places his/her hand on a table and between two antennas such that the transmitted OFDM signal passes through the hand of the subject, and is subsequently collected by the receiver. 

The raw data collected by the receiver due to both (CBDM and HBDM) methods consists of channel frequency response (CFR) that is fed to multiple machine learning (ML) classifiers which eventually determine whether a person is hydrated or dehydrated. {\it To the best of our knowledge, this is the first work that reports a non-contact method for dehydration monitoring.}

{\bf Rationale.}
The proposed CBDM and HBDM methods rely upon the following to infer dehydration related information from the data collected off the chest and the hand of the subject:
i) Dehydration results in reduced blood volume and increased blood viscosity which in turn increases the heart rate and lessens the force of the blood against the walls of the arteries. ii) OFDM signal, being a wideband signal, helps in sensing for dehydration. That is, each OFDM subcarrier captures unique signatures of dehydration due to frequency, phase and amplitude modulation of the subcarrier reflected off the human body. Both factors assist our ML classifiers in achieving high classification accuracy.




{\bf Outline.} 
The rest of this paper is organized as follows. Section II discusses the related work. Section III provides a compact discussion of the apparatus/equipment that provides the scaffolding for our proposed non-contact dehydration monitoring method. Section IV provides further details about the software and hardware setup used for data collection, specifics of each of the two proposed experiments (chest-based, and hand-based), as well as the data acquisition protocol implemented in order to construct our custom  HCDDM-RF-5 dataset. Section V talks about the training and testing of various ML classifiers on our custom dataset, and provides a detailed performance analysis. Section VI concludes. 

\section{Related Work}

The literature on dehydration monitoring is scarce, but could be broadly classified into three categories: i) invasive methods, ii) non-invasive but contact-based methods, iii) non-contact methods. First kind of methods (i.e., invasive methods) which examine blood or urine samples in order to determine the plasma and urine osmolality (and are considered as gold standard) have already been discussed in section I. Further, to the best of our knowledge, there exists no work for third kind of methods (i.e., non-contact methods) for dehydration monitoring in the open literature. Therefore, we summarize the related work on second kind of methods (i.e., non-invasive methods) only. 


\subsection{Non-invasive methods for dehydration monitoring}

The non-invasive methods for dehydration monitoring typically employ wearable sensors (e.g., oximeters, smart watch, smart wrist-bands, etc.) that capture the photoplethysmography (PPG) and electrodermal activity (EDA) signals and pass them through various ML algorithms in order to infer the dehydration status of a subject. 

For example, \cite{posada2019mild} collects both the EDA and the PPG data from 17 subjects and feeds it to a range of ML algorithms in order to detect mild dehydration by exploiting the autonomic response to cognitive stress (induced by means of Stroop test). 
In \cite{liaqat2022personalized}, authors collect EDA data from 16 subjects for three different body postures (sitting, standing, and walking), and pass it to a hybrid Bi-LSTM neural network in order to classify the hydration level of a subject into one of the three different states (hydrated, moderate dehydration, extreme dehydration). 
Authors of \cite{reljin2018automatic} utilize a miniature pulse oximeter to collect PPG data from 17 dehydrated patients admitted in emergency of a tertiary care hospital. They then extract multiple features from the acquired PPG data using the variable frequency complex demodulation algorithm, feed them to a support vector machine classifier, and report an accuracy of $67.91\%$. 
\cite{suryadevara2015towards} collects the EDA data, skin temperature, heart rate and body mass index from 16 participants while they undergo a workout/physical activity known as circuit training. It then feeds this data to an empirically derived formula in order to quantify fluid loss (dehydration) caused by physical activity. 
In \cite{kulkarni2021non}, authors developed a real-time Android-based tool called "monitoring my dehydration" that utilizes the EDA data to learn the dehydration level of a person using machine learning techniques. They did experimental evaluation of their tool by feeding it real-world data from five users, obtaining an accuracy of $84.5\%$.
In \cite{liaqat2020non}, authors collect EDA data using BITalino kit from 5 subjects for three different activities by the subjects (sitting, standing, laying down), feed their data to various ML classifiers to solve the binary classification problem of dehydration detection, and report best classification accuracy of 91.3\% using the random forests ML classifer.
In \cite{rizwan2020non} authors collect EDA data from several subjects under different conditions (sitting, standing), feed it to several ML classifiers to solve the binary classification problem of dehydration detection, and report a maximum accuracy of 87.78\% using the simple k-NN classifier.
Finally, \cite{carrieri2020explainable} takes a rather different approach, and utilizes a leg skin microbiome data from 63 female subjects in order to accurately predict their skin hydration levels and several other important bio-markers.


Before we conclude this section, it is imperative to have a quick discussion about the rise of non-contact methods for remote health sensing in the post-covid19 era.

\subsection{Non-contact methods for health sensing}

The non-contact methods for monitoring of body vitals gained popularity in the post-covid19 era when it was learned that the covid19 pathogen/virus could stay alive on various surfaces for longer duration, and thus, could infect a healthy individual upon contact \cite{hafeez2020review}. This gave rise to non-contact methods which can monitor a person's vital signs from a distance, and thus, could be used for long-term and real-time monitoring of a subject without inconvenience \cite{yatani2012bodyscope,ertin2011autosense}. Such methods also have the potential to decrease the number of visits to a hospital by a patient, thereby reducing the burden on healthcare systems \cite{saeed2021wireless}.

Non-contact health sensing methods could be categorized into following four categories. 

1) Camera-based sensing: These methods begin by recording the face and chest video of a subject from a distance and calculate vital signs by using the periodic change in skin colour to calculate the various body vitals \cite{sato2006non,al2017monitoring}. 

2) Radar-based sensing: These systems incorporate various kinds of radars, e.g., ultra-wideband pulse radar, frequency modulated continuous-wave radar, etc. that utilize the traditional radar principles of range and Doppler in order to estimate various body vitals \cite{van2016wireless}, \cite{massaroni2020contactless}.

3) Wi-Fi-based sensing: Such methods exploit the extensive existing infrastructure of WiFi routers indoors to run cutting-edge ML and deep learning (DL) algorithms on the data collected off the reflections from the human subjects in order to measure body vitals \cite{ali2020goodness}, \cite{massaroni2020contactless}. 

4) Software-defined radio (SDR)-based sensing: Such methods capitalize on the amplitude and phase fluctuations in the signals reflected off the human body to measure vitals \cite{rehman2021rf,pervez2023hand}. 

{\it Note that the proposed non-contact CBDM method and HBDM method both do SDR-based sensing for dehydration monitoring. However, to the best of authors' knowledge, non-contact monitoring of dehydration has not been reported in the open literature, to date.}

\section{Proposed Apparatus for Non-Contact Dehydration Monitoring}
\label{sec:sys-model}
The proposed non-contact system for dehydration monitoring is basically an RF transceiver that consists of two workstations, each connected with a software-defined radio (SDR) by means of a USB 3.0 port (see Fig. $1$). Specifically, the SDR devices used for experiments are Universal Software Radio Peripheral (USRP) model B210\footnote{The USRP B210 from National Instruments covers a wide frequency range (70 MHz to 6 GHz). It can process a wideband spectrum of up to 56 MHz in real time and sample at a high rate of up to 61.44 MS/s.}. Each SDR communicates with other by means of a directional horn antenna. We use MATLAB R2021a to program both the transmit and receive USRP SDRs. Specifically, the transmit SDR sends an orthogonal frequency division multiplexing (OFDM) signal with quadrature phase shift keying (QPSK) modulation on each sub-carrier, while the receive SDR receives it and processes it.

Next, with the aim of non-contact dehydration monitoring, we design two distinct experiments. During the first experiment, the subject's chest is exposed to the OFDM signals, and thus, the receive SDR collects the signal reflected off the chest of the subject. We name this method as chest-based dehydration monitoring (CBDM) method. During the second experiment, the subject's hand is exposed to the OFDM signals, and thus, the receive SDR collects the signal that passes through the hand of the subject. We name this method as hand-based dehydration monitoring (HBDM) method\footnote{This study was approved by the ethical institutional review board (EIRB) of Information Technology University, Lahore, Pakistan.}. 


\begin{figure}[ht]
\begin{center}
	\includegraphics[width=9cm,height=6cm]{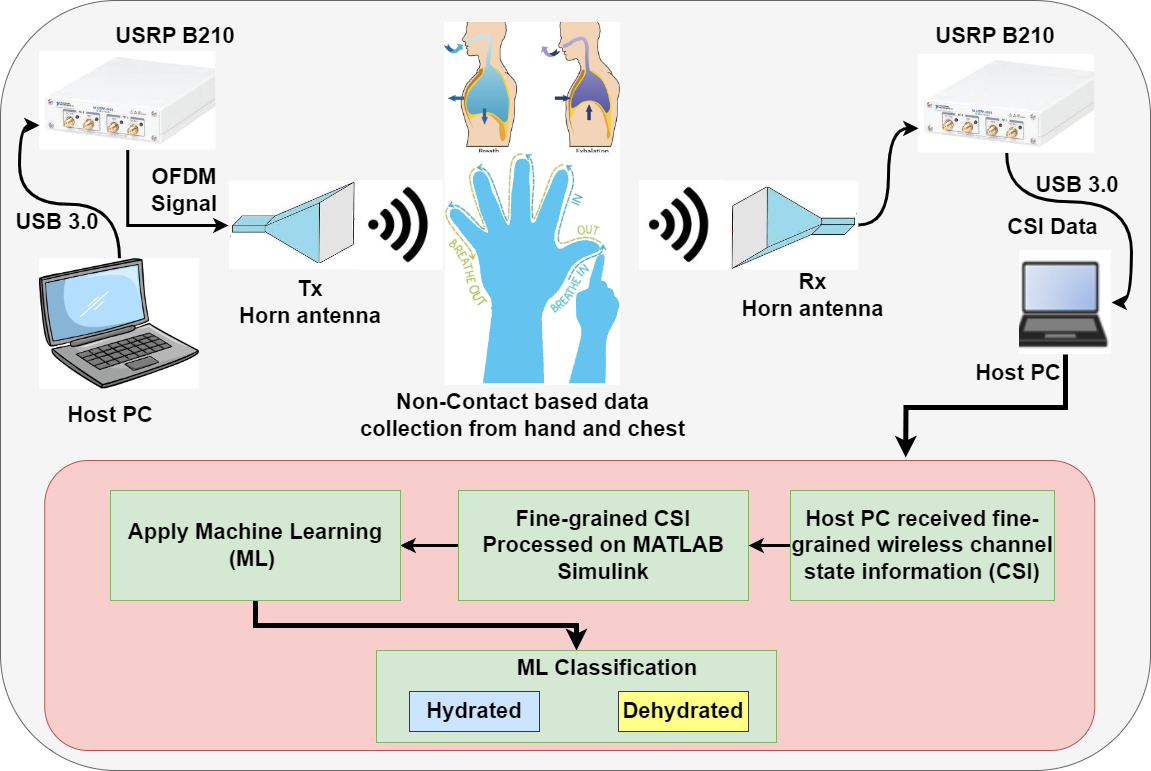} 
\caption{The proposed non-contact method for dehydration monitoring: The apparatus consists of an SDR-based RF transceiver to collect radio data off the chest and the hand of the subject. The collected data is subsequently passed to various machine learning methods, which ultimately classify a subject either as hydrated or dehydrated.}
\label{fig:sysmodel}
\end{center}
\end{figure}




\section{The HCDDM-RF-5 dataset}
\label{sec:method}

This section provides sufficient details about the hardware and software setup used to construct the custom HCDDM-RF-5 dataset\footnote{The acronym HCDDM-RF-5 stands for {\bf H}and and {\bf C}hest {\bf D}ata for {\bf D}ehydration {\bf M}onitoring via {\bf R}adio {\bf F}requency data collected from {\bf 5} subjects.}, our thoughtful data collection methodology (that helped us capture dehydration related data in a very controlled manner), as well as the intricate details of the two experiments performed in order to collect data for the two proposed (CBDM and HBDM) methods. 







\subsection{USRP SDRs based OFDM transceiver}



OFDM Transmitter: 
For each OFDM frame, the random bits generator block creates pseudo-random data bits with a chunk size of 128 bits. The QPSK modulator block maps these bits to (frequency domain) symbols which are then transformed into a time-domain signal by means of an inverse fast Fourier transform (IFFT) of size $N=64$ points. Further, a cyclic prefix (CP) of size 16 samples is appended to each OFDM frame, making each OFDM frame 80 samples long. Gain of the transmit horn antenna is set to 40 dB. Fig. 2(a) shows the Simulink flowgraph of USRP SDR based OFDM transmitter.

OFDM Receiver: 
After removing the CP from each OFDM frame, fast Fourier transform (FFT) is then used to transform the received time-domain OFDM samples into the equivalent frequency-domain OFDM symbol. Then, keeping in mind that the transmitted QPSK symbols on each sub-carrier are known to the OFDM receiver, the channel coefficient $h_i$ for $i$-th sub-carrier could simply be computed as: $h_i=\frac{y_i}{x_i}$, where $x_i$,$y_i$ are the transmitted and received QPSK symbol on $i$-th sub-carrier, respectively. This way, the raw CFR data $\mathbf{h}=[h_1,\cdots,h_N]^T$ is collected by the OFDM transmitter, which is to be utilized later by the ML algorithms in order to classify the status of each subject as either hydrated or dehydrated.
Fig. 2(b) shows the Simulink flowgraph of USRP SDR based OFDM transmitter.


{Table \ref{table:1} provides a quick summary of setting of various relevant parameters of transmit and receive USRP SDRs.}

\begin{figure*}
\hfill
\subfigure[The transmitter flowgraph]{\includegraphics[scale=0.42]{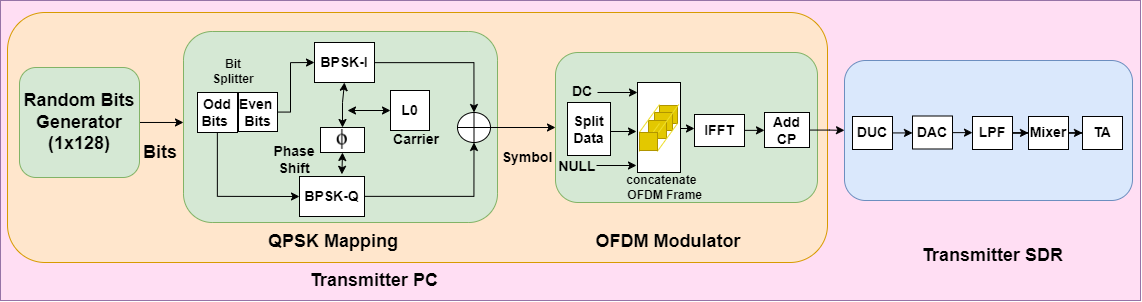}}
\hfill
\subfigure[The receiver flowgraph]{\includegraphics[scale=0.42]{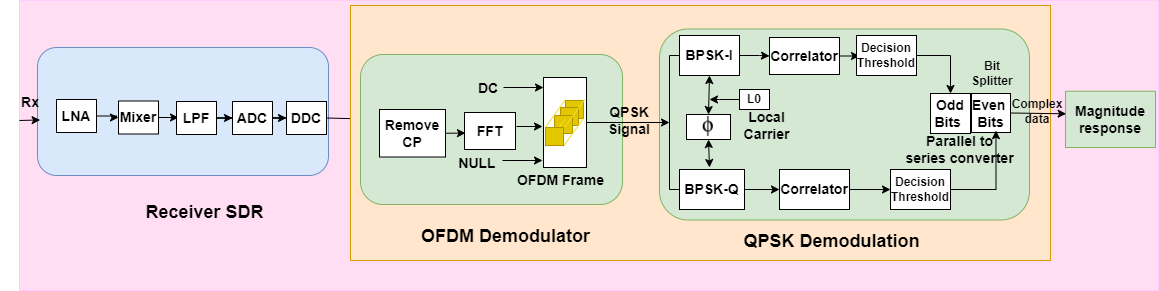}}
\hfill
\caption{The Simulink flowcharts of the USRP-SDR based OFDM transmitter and receiver. }
\label{fig:flowg}
\end{figure*}

\begin{table}[h!]
\centering
\begin{tabular}{|c| c|} 
 \hline
 
 {Parameter} & {Type/Value}
	
\\
\hline

Bits per OFDM frame &	$128$ \\
Bits per symbol &	$2$ \\
Coding scheme &	Gray coding
 \\
Modulation scheme &	QPSK \\
No. of OFDM subcarriers (N) &	$64$\\
Data subcarriers &	$52$\\
Pilot subcarriers &	$12$\\
Size of FFT/IFFT &	$64$ points\\
Size of cyclic prefix	& $16$\\
Sampling rate&	$20,000$ samples/sec\\
Antenna type&	directional horn \\
USRP B210 frequency range&	$70$ MHz - $6$ GHz\\
Centre frequency&	$5.23$ GHz\\
Clock source \& PPS source &	Internal\\
Internal clock rate &	$200$ MHz\\
Interpolation factor (at Tx) &	$250$\\
Decimation factor (at Rx) &	$250$\\
Transmitter gain (at Tx and Rx) &	$40$ dB\\

 \hline
\end{tabular}
\caption{Some parameters for the USRP-SDR-based OFDM transceiver used for non-contact monitoring of dehydration.}
\label{table:1}
\end{table}

\subsection{Data Acquisition for the HCDDM-RF-5 dataset}
The custom HCDDM-RF-5 dataset was constructed by collecting data from five volunteers during the month of Ramadan (between March 23rd, 2023 and April 21st, 2023). Ramadan is an Islamic holy month during which devout Muslims observe a strict fast from sunrise till sunset. That is, while they are fasting, Muslims refrain from eating and drinking from sunrise till sunset. We took advantage of this unique opportunity in order to collect dehydration related data from five devout Muslims who had been fasting during this month. Among five subjects, two were males (aged 28, 62 years), and three were females (aged 21, 26, 61 years). For each fasting subject, we collected data twice, once for each class label (hydrated and dehydrated) in order to construct a balanced dataset. Specifically, first episode of data collection took place about 30 minutes before the sunset when the subject was deemed to be maximally dehydrated (thus, this data belongs to the first/dehydrated class). Subsequently, the second episode of the data collection took place an hour after the sunset, after the subject had finished eating and drinking after breaking the fast (thus, this data belongs to the second/hydrated class). For each subject, we conducted two kinds of experiments where we exposed the subject's chest and hand to the RF signals, respectively. Some more pertinent details about data collection for our proposed CBDM and HBDM methods are given below. 



{\it Data collection for the proposed CBDM method:} 
During data acquisition for the proposed CBDM method, each participant sat on a chair that was about 80 cm away from the pair of directional horn antennas that pointed towards the chest of the subject (see Fig. 3). As described before, the transmit horn antenna impinged an OFDM signal onto the chest of the subject, while the receive horn antenna gathered the signal reflected off the subject's chest. During each experiment session, the subject sat still in order to avoid motion-induced artefacts in the data being gathered. Each single experiment session lasted for 30 seconds. For each subject, we conducted five experiment sessions before the sunset (to capture the raw CFR data for dehydrated class) and five experiment sessions after the sunset (to capture the raw CFR data for the hydrated class). This way, we were able to collect $30\times5=150$ seconds worth of data for each class (for a given subject), and thus, $150\times2=300$ seconds worth of data per subject. Ultimately, for 5 subjects, this led to a total dataset size of $300\times5=1500$ seconds (or, 25 minutes) of raw CFR data (that corresponds to a total of $5\times5\times2=50$ experiment sessions). 

\begin{figure}[ht]
\begin{center}
	\includegraphics[width=8cm,height=4cm]{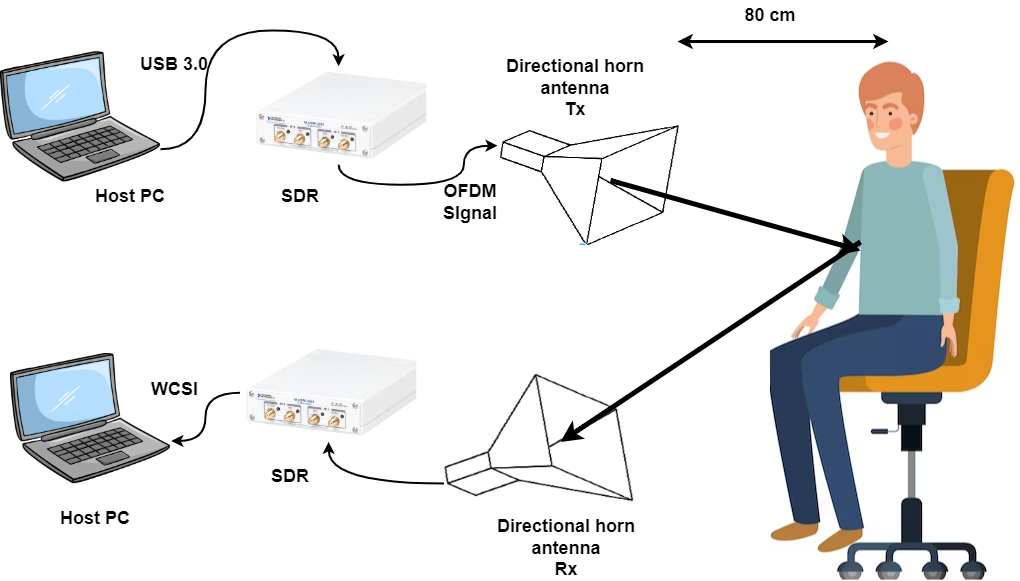} 
\caption{Experimental setup of the proposed CBDM method.}
\label{fig:method1}
\end{center}
\end{figure}

{\it Data collection for the proposed HBDM method:} 
During data acquisition for the proposed HBDM method, each participant sat on a chair that was about 60 cm away from the pair of directional horn antennas facing each other, and placed his/her hand on the table between the two antennas (see Fig. 4). Again, the transmit horn antenna impinged an OFDM signal onto the hand of the subject, while the receive horn antenna gathered the signal passed through the subject's hand. During each experiment session, the subject sat still in order to avoid motion-induced artefacts in the data being gathered. The rest of the details of data acquisition for the proposed HBDM method are the same as before. That is, for each subject, we conducted five experiment sessions before the sunset (to capture the raw CFR data for dehydrated class) and five experiment sessions after the sunset (to capture the raw CFR data for the hydrated class). This way, for 5 subjects, we acquired a dataset that consisted of $300\times5=1500$ seconds (or, 25 minutes) of raw CFR data (that corresponds to a total of $5\times5\times2=50$ experiment sessions). 

In short, combining the two smaller datasets due to CBDM method and HBDM method together, the custom HCDDM-RF-5 dataset consists of a total of 50 minutes of raw CFR data that corresponds to a total of 100 experiment sessions.


\begin{figure}[ht]
\begin{center}
	\includegraphics[width=9cm,height=3cm]{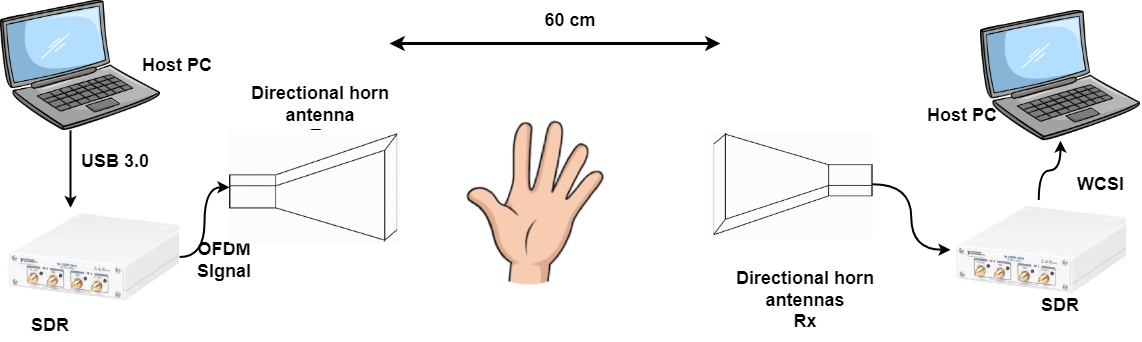} 
\caption{Experimental setup of the proposed HBDM method. }
\label{fig:methodour}
\end{center}
\end{figure}

\section{Training and Testing of Machine Learning Classifiers}

For the binary classification problem (hydrated/dehydrated) under consideration, we train and test the following five ML classifiers and their variants: K-nearest neighbours (KNN), support vector machine (SVM), decision tree (DT), ensemble classifier, and neural network. 
Subsequently, we provide detailed performance analysis and comparison of all the ML classifiers implemented. 

\subsection{Data Pre-processing \& Training of Machine Learning Classifiers }  

{\it Data Pre-processing:}
We utilised a low-pass filter and a Savitzky-Golay filter to denoise the CFR extracted from the received OFDM signal, for all the experiment sessions (for both CBDM and HBDM methods). We inspected the whole data manually and removed artifacts where found.

{\it Training \& validation of ML classifiers: }
The Matlab's classification learner app was used to train the following ML classifiers:
K-nearest neighbour (KNN), support vector machine (SVM), decision tree (DT), ensemble classifier, and neural network. All the classifiers were trained on both labelled datasets (corresponding to the CBDM method and the HBDM method). The K-fold cross-validation strategy was used for validation in order to prevent the over-fitting issue. 


\subsection{Performance metrics}

Each classifier's performance is quantified in terms of accuracy, given as: 
\begin{align}
   & \mathrm{Accuracy}=\frac{\mathrm{Correct \; prediction}}{\mathrm{Total \; observations}}\times 100
\\
&\mathrm{Accuracy}=\frac{T_n+T_p}{T_n+T_p+F_n+F_p} \times 100  
\end{align}
where $T_n$ represents a true negative, $T_p$ represents a true positive, $F_n$ represents a false negative, and $F_p$ represents a false positive. In addition, we also do a performance comparison of the various ML algorithms by means of a confusion matrix.


\subsection{Performance of proposed CBDM method }

We begin with performance analysis of the k-NN classifier for three distinct values of k, i.e., k=1,k=10,k=100 (where k is the number of neighbours used to calculate the distance to the new data point). We learn that the fine k-NN (k=1) achieves an accuracy of 79.1$\%$, medium k-NN (k=10) achieves an accuracy of 69.2$\%$, while the coarse K-NN (k=100) achieves a very low accuracy of 55.3$\%$ (see Fig. 5 that displays the detailed confusion matrix).

Next, we focus on Fig. 6 and do performance comparison of the remaining four ML classifiers (and their variants). Beginning with an SVM classifier (with linear, quadratic, and cubic kernels), we note that the linear SVM achieves an overall accuracy of 86.5\%, quadratic SVM achieves an overall accuracy of 89.6\%, while the cubic SVM achieves an overall accuracy of 90.9\%. 
Next, we focus on the decision tree classifier, and note that it has the lowest accuracy of all. That is, the fine tree (despite its many leaves and despite its ability to differentiate between classes precisely) achieved an accuracy of 68.8\% only, while the coarse tree achieved a very low accuracy of 58.0\% only. 
Next in line is the ensemble classifier (a mixture of many classifiers) that is typically implemented with the aim to boost classification accuracy. We observe the following: the ensemble boosted tree has an overall accuracy of 70.3\%, the ensemble bagged tree has an accuracy of 77.9\%, the ensemble subspace KNN has an accuracy of 82.9\%, while the ensemble subspace discriminant has an accuracy of 89.6\%. 
Finally, the neural network (NN) classifier. Each variant of the NN classifier is a fully-connected feedforward network. After each fully connected layer, the Relu activation function is applied, except the last year where softmax activation function is used. We observe that all the different variants of the NN classifier outperform the other ML classifiers. Specifically, the narrow variant of the neural network achieves an accuracy of 93.8\%, the medium neural network achieves an accuracy of 92.5\%, the broad neural network achieves an accuracy of 92.9\%, the bi-layered variant of neural network achieves an accuracy of 93\%, while the tri-layered variant of the neural network achieves an accuracy of 93.1\%. 

Fig. $7$ provides an alternate way of comparing the overall accuracy of all the five ML classifiers and their variants. We note that, for the proposed CBDM method, the neural network classifier (with the narrow neural network) achieves the highest accuracy, which is 93.8\%.

\begin{figure}[t]
\begin{center}
	\includegraphics[scale=0.3]{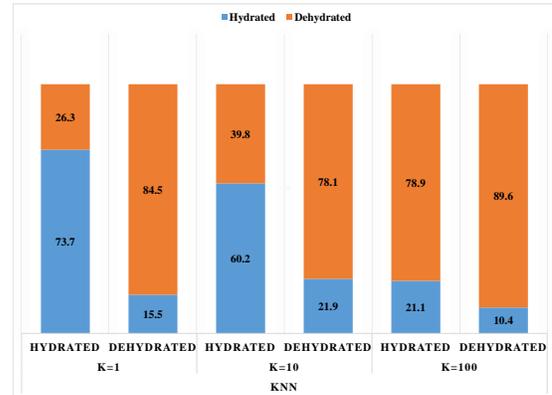} 
\caption{Confusion matrix of k-NN algorithm for the proposed CBDM method.}
\label{fig:methodour}
\end{center}
\end{figure}

\begin{figure}[t]
\begin{center}
	\includegraphics[scale=0.35]{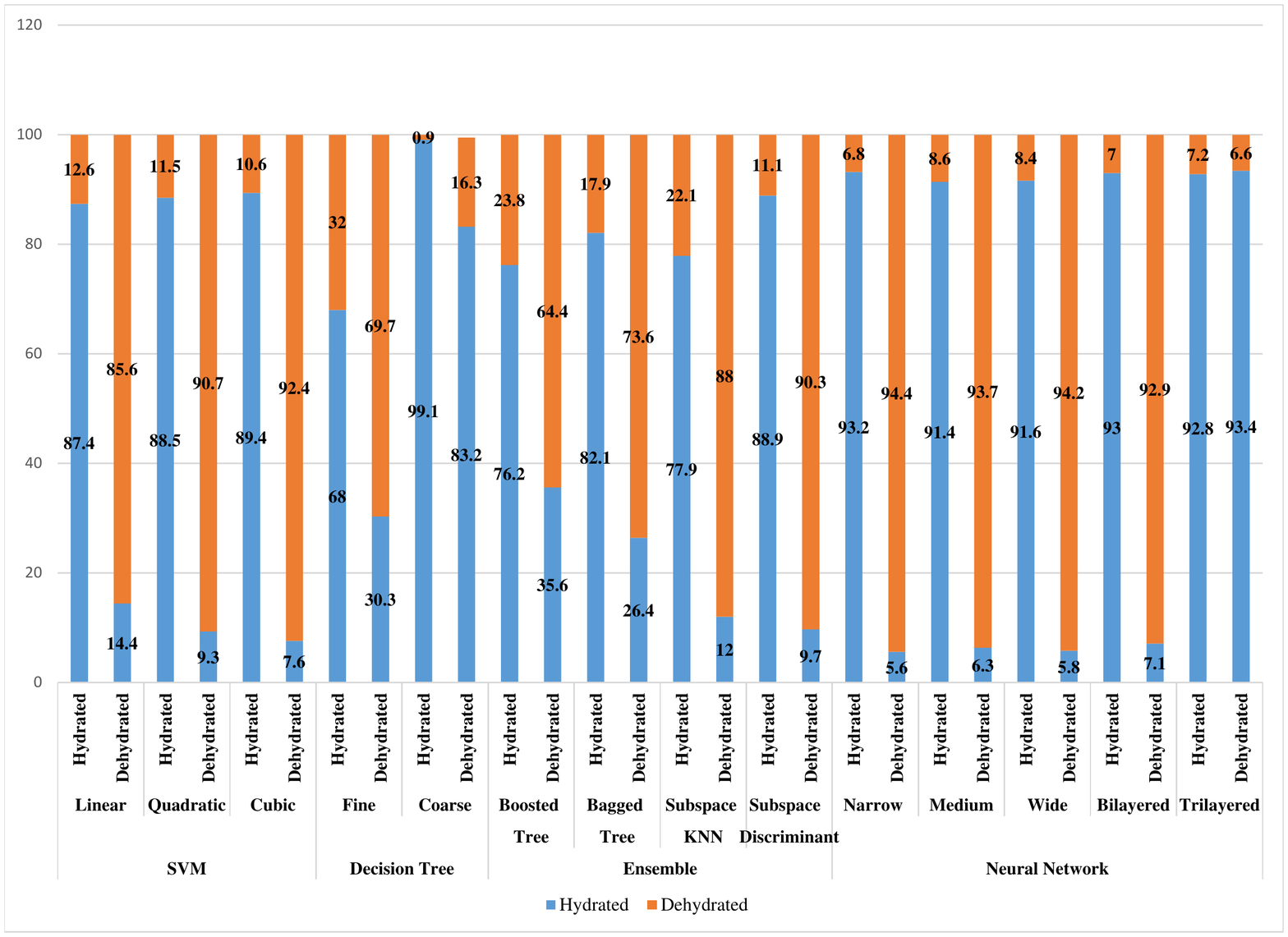} 
\caption{Confusion matrix of each of SVM, DT, Ensemble classifiers, and NN for the proposed CBDM method.}
\label{fig:methodour}
\end{center}
\end{figure}

\begin{figure}[ht]
\begin{center}
	\includegraphics[scale=0.33]{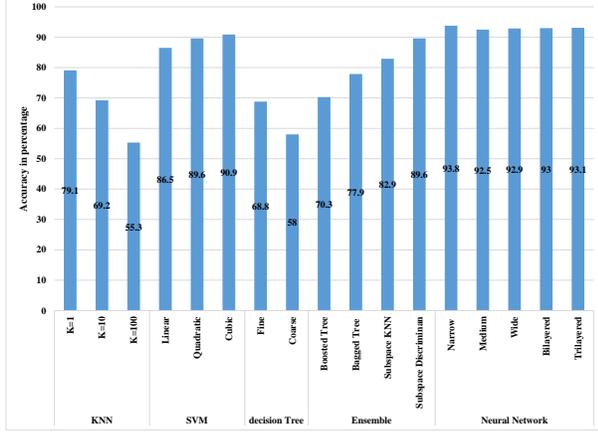} 
\caption{Performance comparison of all the classifiers for the proposed CBDM method.}
\label{fig:methodour}
\end{center}
\end{figure}

\subsection{Performance of proposed HBDM method}

We begin performance analysis of our proposed HBDM method from Fig. 8 which provides confusion matrix of each of five ML classifiers (and their variants). 
Beginning with an SVM classifier (with linear, quadratic, and cubic kernels), we note that the linear SVM achieves an overall accuracy of 71.1\%, quadratic SVM achieves an overall accuracy of 89.2\%, while the cubic SVM achieves an overall accuracy of 88.2\%. 
Next, the decision tree classifier. We observe that once again it has the lowest accuracy of all. That is, the fine tree achieved an accuracy of 72.2\% only, while the coarse tree achieved a very low accuracy of 61.4\% only. 
Next, the ensemble classifier. We observe the following: the ensemble boosted tree has an overall accuracy of 74.8\%, while the ensemble bagged tree has an accuracy of 79.7\%. 
Finally, the neural network (NN) classifier. Once again, all the different variants of the NN classifier outperform the other ML classifiers. Specifically, the narrow variant of the neural network achieves an accuracy of 94.7\%, the medium neural network achieves an accuracy of 96.15\%, the broad neural network achieves an accuracy of 95.15\%, the bi-layered variant of neural network achieves an accuracy of 92.35\%, while the tri-layered variant of the neural network achieves an accuracy of 94.2\%. 

Fig. 9 provides an alternate way of comparing the overall accuracy of all the five ML classifiers and their variants. We note that, for the proposed HBDM method, the neural network classifier (with the medium neural network) achieves the highest accuracy, which is 96.15\%.

\subsection{Performance comparison with the state-of-the-art}
Finally, Table II compares the accuracy of the proposed non-contact CBDM and HBDM methods with the state-of-the-art methods which are all contact-based methods for dehydration monitoring. Compared to the state-of-the-art where the maximum reported accuracy is 97.83\%, our proposed non-contact method is slightly inferior (as we report a maximum accuracy of 96.15\%); nevertheless, the advantages of
our non-contact dehydration method speak for themselves. That is, our proposed method is non-invasive and contact-less, has high accuracy, allows continuous and seamless monitoring, is easy to use, and provides rapid results.

\begin{figure}[t]
\begin{center}
	\includegraphics[scale=0.33]{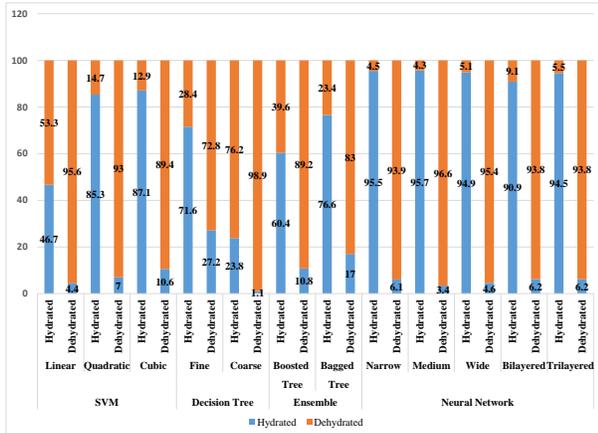} 
\caption{Confusion matrix of each of KNN, SVM, DT, Ensemble classifiers, and NN for the proposed HBDM method.}
\label{fig:methodour}
\end{center}
\end{figure}

\begin{figure}[ht]
\begin{center}
	\includegraphics[scale=0.33]{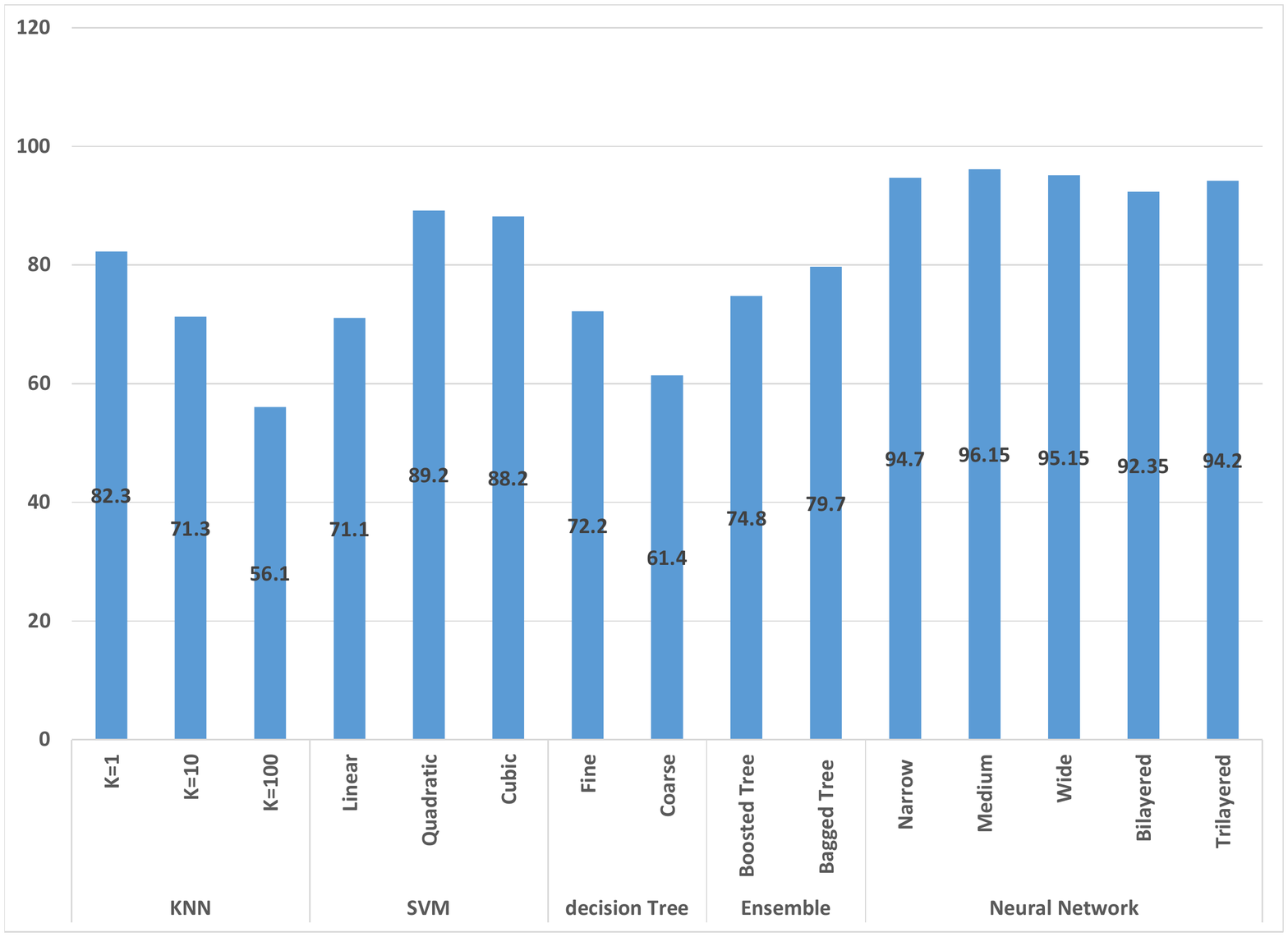} 
\caption{Performance comparison of all the ML classifiers for the proposed HBDM method.}
\label{fig:methodour}
\end{center}
\end{figure}

 \begin{table}[h!]
\centering
\begin{tabular}{|c| c|} 
 \hline
 Work &
	Accuracy
\\
\hline
Liaqat et al. \cite{liaqat2022personalized} & 97.83$\%$ \\
\hline
Kulkarni et al. \cite{kulkarni2021non} & 75.96$\%$ \\
\hline
Liaqat et al. \cite{liaqat2020non} & 91.53$\%$ \\
\hline 
Rizwan et al \cite{rizwan2020non} & 85.63$\%$ \\
\hline 
Carrieri et al. \cite{carrieri2020explainable} & 73.91 $\%$ \\
\hline  
Our non-contact CBDM method & 93.8$\%$ \\
\hline 
Our non-contact HBDM method & 96.15$\%$ \\
\hline 

\end{tabular}
\caption{Accuracy comparison of our proposed non-contact CBDM and HBDM methods with the state-of-the-art (contact-based methods). }
\label{table:propmoverall}
\end{table}



\section{Conclusion \& Future Work}
\label{sec:conclusion}

This work proposed for the first time a non-contact method to monitor the dehydration of a subject from a distance. Specifically, we utilized a pair of USRP SDRs whereby the transmit SDRs impinged OFDM signals onto the chest or the hand of the subject, while the receive SDR collected the modulated signal reflected off the body of the subject. For the purpose of training our ML classifiers, we collected data from 5 Muslim subjects (before and after sunset) who were fasting during the month of Ramadan. We then passed the received raw CFR data through many ML classifiers. Among them, neural network classifier achieved the best performance: an accuracy of 93.8\% for the proposed CBDM method, and an accuracy of 96.15\% for the proposed HBDM method. The fact that the proposed HBDM method outperforms the proposed CBDM method is a pleasant result. This is because this allow us to promote the proposed HBDM method as a non-contact method for dehydration monitoring (where only a hand is exposed to RF radiation, instead of the full chest, albeit the radiation being non-ionizing). Last but not the least, the proposed non-contact method (with a maximum accuracy of 96.15\%) performs very close to its contact-based counterpart (with a maximum accuracy of 97.83\%). Such a minor performance degradation of our proposed non-contact method compared to its contact-based competitor might be affordable, keeping in mind the convenience (and other benefits) of a non-contact method. 

One major advantage of the proposed approach is that it may pave the way for the creation of a smart mobile health (m-health) solution that could be deployed in remote areas far away from the mega cities, in order to provide comprehensive health monitoring of the people living there. 

This work opens up many exciting directions for the future work. For example, one could construct/acquire a more challenging dataset (unlike the current dataset that was obtained in a very controlled setting), and re-evaluate as well as fine-tune the performance of the proposed method further, in order to make it robust and amicable to the unseen data.

\footnotesize{
\bibliographystyle{IEEEtran}
\bibliography{references}
}

\vfill\break

\end{document}